# Vitrification-Devitrification Enables Tunable Photonic and Gas Sorption Properties of Zeolitic Imidazolate Frameworks

Zhencai Li[1], Zihao Wang[2], Minhyuk Kim[3], Huotian Zhang[4], Bozhao Yin[5], Yong Li[1], Qi Zhang[6], Fengming Cao[7], Xuan Ge[7], Laurent Calvez[8], Daniel Irving[9], Guoping Dong[5], Feng Gao[4], Haomiao Zhu[2], Morten M. Smedskjaer[1], Hoi R. Moon[3*], Yuanzheng Yue[1*]

[1]Department of Chemistry and Bioscience, Aalborg University, DK-9220 Aalborg, Denmark
[2]Xiamen Research Center of Rare Earth Materials, Haixi Institutes, Chinese Academy of Sciences, Xiamen 361021, China
[3]Department of Chemistry and Nanoscience, Ewha Womans University, Seoul 03760, Republic of Korea
[4]Department of Physics, Chemistry, and Biology (IFM), Linköping University, Linköping 583 30, Sweden
[5]School of Materials Science and Engineering, South China University of Technology, Guangzhou 510640, China
[6]College of Electronic and Optical Engineering and College of Flexible Electronics (Future Technology), Nanjing University of Posts & Telecommunications, Nanjing 210023, China
[7]Shanghai Key Laboratory of Materials Laser Processing and Modification, School of Materials Science and Engineering, Shanghai Jiao Tong University, 200240 Shanghai, China
[8]Univ Rennes 1, CNRS, ISCR (Institut des Sciences Chimiques de Rennes) - UMR 6226, F-35000, Rennes, France
[9]Diamond Light Source, Harwell Science and Innovation Campus, Oxfordshire OX11 0QX, UK

Corresponding author. Email: hoirimoon@ewha.ac.kr; yy@bio.aau.dk

**Abstract:**

Zeolitic imidazolate framework (ZIF) glasses represent an emerging family of melt-quenched glasses, which exhibit immense potential for applications in gas separation, energy storage, and optics. However, their intrinsic porosity remains elusive due to the inherent challenges in resolving their disordered atomistic structures. Here, we systematically investigate the porosity of ZIF-4 and ZIF-62 crystals and their corresponding glasses. $CO_2$ sorption at 195 K enables quantitative assessment of microporosity in both crystalline and glassy states, allowing the accessible micropore volume of the ZIF glasses to be determined. Moreover, establishing a direct relationship between photonic properties and structural porosity in Zn-based ZIF glasses remains challenging. Here we demonstrate striking broadband blue-light emission from ZIF-4 glass annealed under optimized conditions. A pronounced red shift is observed when increasing the annealing temperature above the glass transition temperature. By correlating the evolution of photoluminescence with structural porosity, we reveal the interplay between the photonic and gas sorption properties of ZIF glasses. These findings provide new insights into the structure-property relationships of ZIF glasses and offer a pathway toward the rational design of multifunctional MOF glasses.

## 1. Introduction

Metal-organic frameworks (MOFs), known as nanoporous crystals, consist of metal ions coordinated to organic ligands and possess both large specific surface areas and tunable pore structures that enable various applications in gas adsorption[1],[2],[3], catalysis[4], and energy storage[5]. Zeolitic imidazolate frameworks (ZIFs), a subclass of MOFs, are constructed through Werner-type metal-ligand coordination[6] and can be transformed into the glass state via melt-quenching. In addition, ZIF glasses largely retain the pore structures of their crystalline precursor[7]. Although more than 250 crystalline ZIFs spanning over 50 distinct network topologies have been reported[8], only a small fraction have been demonstrated to melt and form glasses via melt-quenching. ZIF-4 ($M(Im)_2$) and ZIF-62 ($M(Im)_{2-x}(bIm)_x$) are representative glass-forming ZIFs, where M denotes a metal node such as $Zn^{2+}$ or $Co^{2+}$, and Im and bIm are imidazolate ($C_3H_3N_2^-$) and benzimidazolate ($C_7H_5N_2^-$) linkers, respectively[9][10].

ZIF-62 demonstrates ultrahigh glass-forming ability and a wide temperature window of the molten state between its liquidus and decomposition temperatures[11]. Thus, ZIF-62 can be easily vitrified, e.g., by using the spark plasma sintering (SPS) technique[12]. The compounds Im and BIm play a critical role in organic luminescence due to their outstanding optical absorption and $\pi$-$\pi^*$ electronic transition in the *p*-orbital[13], while crystalline ZIF-62 exhibits luminescent properties that differ from those of its individual linkers[14]. That is, the luminescence behaviour depends on its structure and metal-ligand interactions, such as metal-to-ligand charge transfer (MLCT) and ligand-to-metal charge transfer (LMCT)[15]. However, ZIF-62 crystal only exhibits strong ultraviolet light emission, but not visible light emission for white LEDs, because of its too wide band gap (around 3.5 eV). To obtain visible light emission in ZIF-62, Li et al reported white-light emission, which was achieved by a vitrification-pressurization-annealing strategy. However, the white-light emission mechanism of annealed ZIF-62 glasses remains elusive due to the presence of two distinct emission centers associated with the Im and bIm ligands. Consequently, establishing a direct relationship between the photonic properties and ligand chemistry of Zn-based ZIF glasses remains challenging.

In addition to ZIF-62, ZIF-4 should be a simple case to establish a direct relationship between the photonic properties and ligand chemistry. Upon heating, crystalline ZIF-4 first collapses into a low-density amorphous phase at ~315 °C, subsequently transforms into a high-density amorphous phase, and then recrystallizes into the denser ZIF-zni polymorph at ~460 °C before melting at ~580 °C. Quenching the melt to room temperature yields a melt-quenched ZIF-4 glass[16]. Positron annihilation lifetime spectroscopy (PALS) revealed that melt-quenched ZIF-4 possesses residual microporosity[17]. However, this method cannot determine whether the pores identified in ZIF glasses are accessible to gas molecules, and thus whether they form an open pore network[18]. Therefore, PALS cannot readily provide a quantitative measure of porosity, such as the specific pore volume of ZIF glasses.

In this work, we discovered broad blue light emission in ZIF-4 glasses upon hot-pressing and subsequent annealing, which was much stronger than that in their crystalline counterparts. This function was significantly enhanced by annealing above the glass transition temperature, and simultaneously a red shift was induced. We clarified the structural and electronic origin of both the emission enhancement and the red shift. For the gas sorption, we demonstrate that $CO_2$ gas sorption measurements at 195 K provide deep insights into the intrinsic porosity (pore volume) of ZIF-4 and ZIF-62 glasses. This study provides important insights into the intrinsic porosity of ZIF glasses and advances our understanding of the relationship between structural porosity and photonic properties in these materials.

## 2. Results and Discussion

### 2.1. Thermal and structural analyses

Fig. 1a shows the preparation procedure of the studied samples. The crystalline or amorphous nature of the studied samples was identified from the X-ray diffraction (XRD) and high-energy synchrotron XRD (HEXRD) analyses (Figs. 1b and S1, Supporting Information). ZIF-4 crystals were successfully synthesized by using the modified method based on that reported in ref[19], as confirmed by the good agreement between the experimental XRD pattern and the simulated one. The as-synthesized ZIF-4 crystals were subsequently vitrified to form a high-density amorphous (HDA) glass[7], as evidenced by the disappearance of the ZIF-4 diffraction peaks (Fig. 1b). HDA glass was subjected to hot-pressing using spark plasma sintering (SPS), resulting in a hot-pressed high-density amorphous (HPHDA) glass. Interestingly, several weak XRD diffraction peaks emerge on the broad amorphous hump of HPHDA (Figs. 1b and S2, Supporting Information), suggesting the formation of tiny ZIF-zni crystals in the glass matrix. To study the impact of annealing on the growth of ZIF-zni crystals, HPHDA was annealed under different conditions to obtain $A_{xxx\text{-}yy}$HPHDA, where xxx and yy denote the annealing temperature (in K) and annealing time (in min), respectively. In this procedure, HPHDA was annealed at 593 K for 30 min, yielding $A_{593\text{-}30}$HPHDA (see Methods). The HEXRD patterns of $A_{300\text{-}10}$HPHDA and $A_{360\text{-}10}$HPHDA prove that the crystallinity of HPHDA increases with annealing temperatures (Supplementary Fig. 3b).

To characterize the morphologies of tiny ZIF-zni crystals in the glass matrix, transmission electron microscopy (TEM) analysis was conducted on HPHDA and $A_{613\text{-}10}$HPHDA. Figs. 1c and S3 (Supporting Information) show TEM and high-resolution TEM (HRTEM) images of HPHDA. Unlike HDA glass, the HRTEM image of HPHDA reveals abundant crystal-like clusters or paracrystallites (shown in white dashed circles) with sizes of around 2-3 nm, which can be assigned to the ZIF-zni phase. This observation is consistent with the XRD results of HPHDA, but the selected area electron diffraction (SAED) pattern shows no diffraction spots. Compared with the diffuse diffraction rings observed in the SAED pattern of HDA glass, the well-defined diffraction rings $A_{613\text{-}10}$HPHDA suggest a higher degree of crystalline ordering (Figs. 1d and S4, Supporting Information). The ZIF-zni crystals grow to approximately 6 nm. In addition, the SAED photograph (Fig. S5, Supporting Information) confirms the identity of the forming crystals. Energy dispersive X-ray spectroscopy (EDS) mapping analysis, as collected from the high-angle annular dark-field imaging scanning TEM (HAADF-STEM) images, confirms the homogeneous distribution of Zn, C, and N elements in the samples (Fig. S6, Supporting Information).

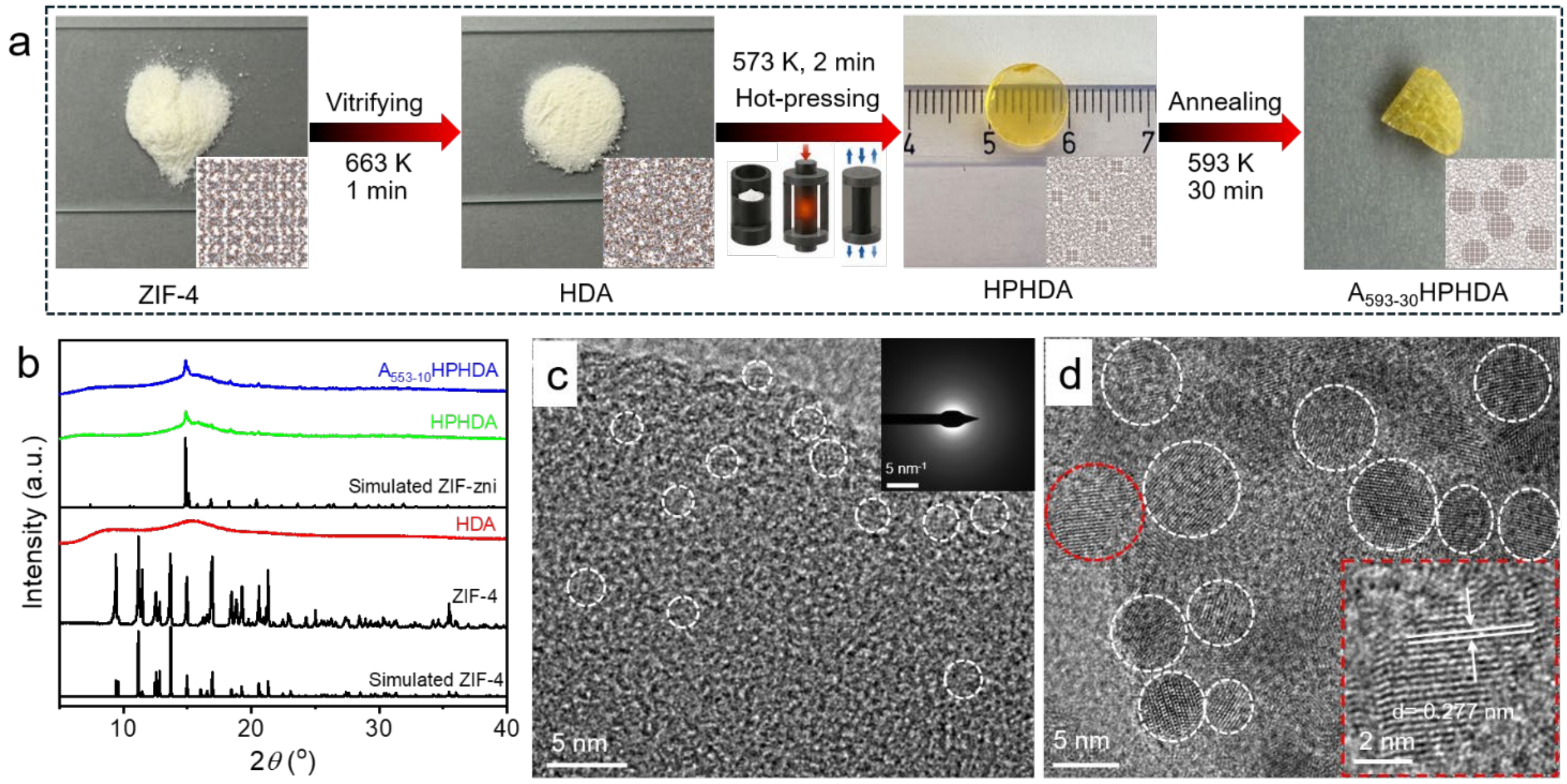

**Fig. 1 Synthesis and characterization of ZIF-4 crystal and glasses.** (a) Flow chart of high-density amorphization (HDA) glass applied to ZIF-4 crystal, hot-pressing (HP) applied to HDA glass powder, and annealing applied to hot-pressed HDA (HPHDA) glass. (b) XRD patterns of simulated ZIF-4, ZIF-4 crystal, HDA glass, simulated ZIF-zni, HPHDA glass, and $A_{553-10}$HPHDA glass. (c-d) Transmission electron microscopy (TEM) images of HPHDA glass (c) and $A_{613-10}$HPHDA glass (d).

To investigate the thermodynamic features of ZIF-4, HDA, HPHDA, and $A_{613-10}$HPHDA, the differential scanning calorimetry (DSC) analysis was performed, and the obtained DSC curves are shown in Fig. 2a. The endotherm (between 475 and 575 K) in the DSC upscan curve of ZIF-4 is attributed to the desolvation of DMF, which does not cause framework collapse[7]. Upon further heating to 663 K, the crystalline ZIF-4 transforms into a low-density liquid (LDL) phase and further converts to a high-density liquid (HDL) phase, without recrystallization into the dense ZIF-zni structure. After completion of the liquid-liquid transition (LLT), cooling the HDL phase to room temperature results in the formation of HDA glass, which exhibits a glass transition temperature ($T_g$) of 564 K upon reheating. Upon hot-pressing (HP), the $T_g$ of HDA glass decreases from 564 K to around 523 K in HPHDA. This decrease was attributed to the pressure-induced weakening of the intermediate-range structural connectivity[12][14]. Upon annealing at 613 K (90 K above $T_g$ of HPHDA) for 10 min, $T_g$ of HPHDA increases to around 563 K, which is similar to that of HDA glass. To study the impact of annealing on the increase in $T_g$ value, the HPHDA sample was subjected to 5 cycles of DSC scan up to the maximum scanning temperature (613 K) at 10 K min$^{-1}$. It is seen that $T_g$ first sharply increases and then slowly increases with cycling numbers, and approaches 563 K in the fifth upscan. Finally, the $T_g$ of the HPHDA sample remains unchanged upon increasing annealing temperatures (633, 653, and 673 K) for 10 min (Fig. S7, Supporting Information). This implies that annealing relaxes the hot-pressed-induced distorted structure and increases glass network connectivity, allowing the structure to return toward that of HDA glass. This phenomenon could also be attributed to the partial crystallization of ZIF-zni, in contrast to the annealing-induced structural evolution observed in hot-pressed ZIF-62 glass[14].

To understand the structure changes caused by vitrification, hot-pressing, and annealing, we analyze the Raman spectra of ZIF-4, HDA, HPHDA, and $A_{673\text{-}30}$HPHDA samples (Fig. 2c). The peaks in the range of 600-1600 cm$^{-1}$ at 1065 cm$^{-1}$ are assigned to the different vibration modes of the Im ring for the four samples[20]. These vibration bands occur almost at the same frequencies, confirming the integrity of organic linkers in the samples[11]. However, compared to the vibration peaks of the Im ring in ZIF-4, those in the other three samples become weaker and broader. In the low-frequency range (below 600 cm$^{-1}$), the vibration bands in ZIF-4 were observed at 145 and 179 cm$^{-1}$ corresponding to Zn-N stretching. Two peaks weaken and merge to form a broad Zn-N peak upon vitrification, hot-pressing, and annealing, implying that vitrification distorts [$ZnIm_4$] tetrahedra and consequently leads to structural disordering in different length scales.

The HEXRD technique is further performed to probe the short-, medium-, and long-range structural evolution of the ZIF-4 crystal upon vitrification, hot-pressing, and annealing. Fig. 2d shows the Faber-Ziman structure factor $S(Q)$ curve of ZIF-4, HDA, HPHDA, and $A_{633-10}$HPHDA, where sharp diffraction peaks are present in ZIF-4 and broad peaks appear in HDA glass. The glassy nature of HDA is further verified by the lack of sharp diffraction peaks in the $S$(Q) curves. Some weak diffraction peaks emerge in the $S$(Q) curves of HDA upon hot-pressing and become more pronounced with increasing annealing temperature, further confirming the occurrence of a disorder-to-order transition during both hot-pressing and annealing. The reduced pair correlation functions $G(r)$ curves (Fig. 2e) show five main atom-atom correlation peaks in the four samples. The atom-atom pair correlations include those of C-C(N) (1.36 Å), Zn-N (2.02 Å), Zn-C (3.03 Å), Zn-N (4.17 Å), and Zn-Zn (5.94 Å), as shown in Fig. 2f. The local coordination environments of $Zn^{2+}$ in the HDA, HPHDA, and $A_{633-10}$HPHDA samples slightly differ from those in ZIF-4. The Zn-Zn correlation of 5.94 Å suggests the integrity of Zn-ligand-Zn linkages, and the correlation in the glasses is weaker than that in ZIF-4, confirming that the three glasses possess a higher degree of intermediate-range disorder. However, as shown in the $\Delta G(r)$ curves (Fig. S8, Supporting Information), the short-range correlations of Zn-N (2.02 Å), Zn-C (3.03 Å), and Zn-N (4.17 Å) become more evident upon hot-pressing and annealing, implying that the coordination Zn-N bonds get stronger.

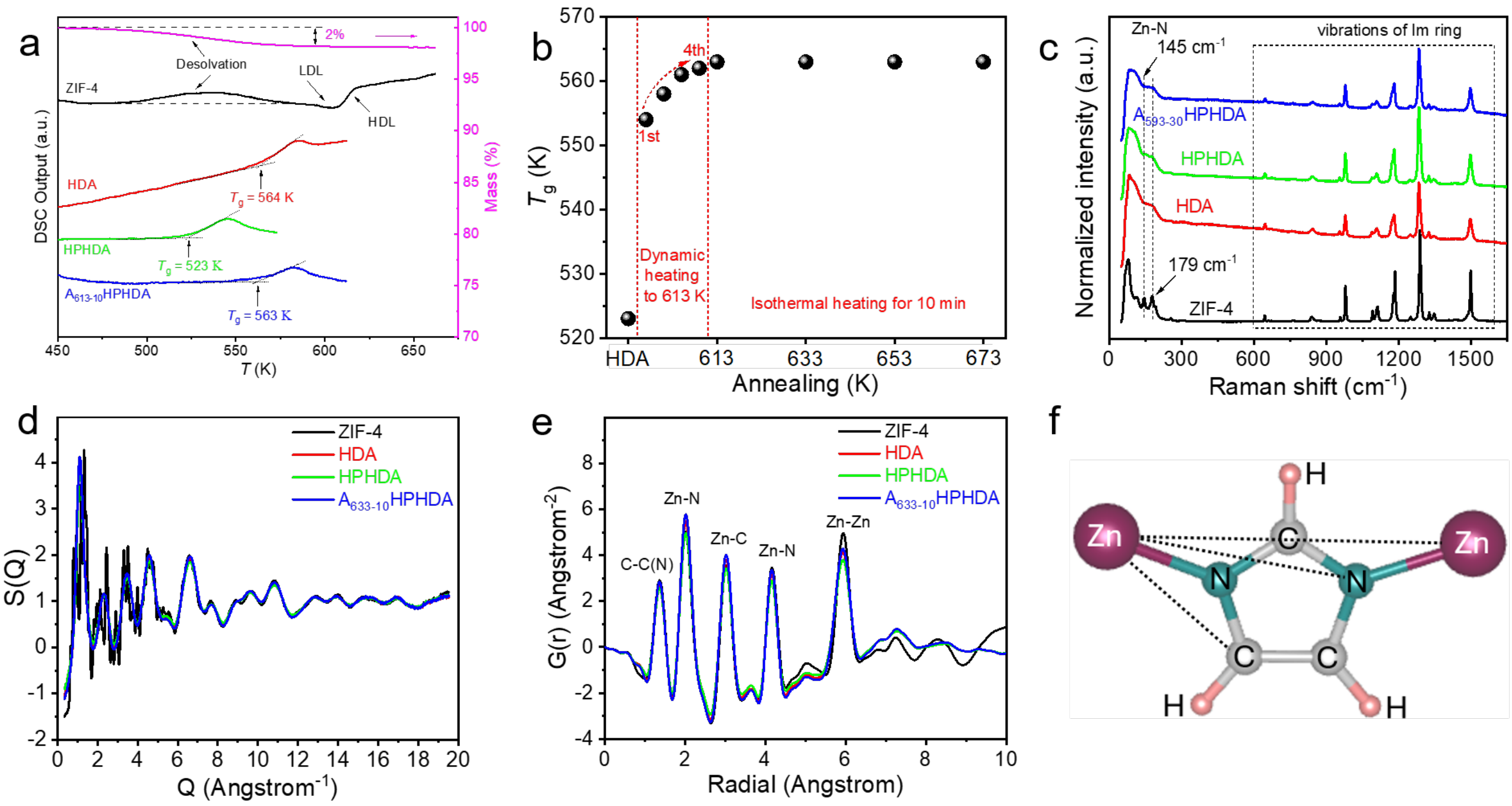

**Fig. 2 Structural analyses of ZIF-4 crystal and glasses.** (a) Thermogravimetric curve (magenta color) of ZIF-4 and differential scanning calorimetry (DSC) curves of ZIF-4, HDA glass, HPHDA, and $A_{613\text{-}10}$HPHDA. (b) The variation of glass transition temperature ($T_g$) for HDA-g with annealing via both dynamic heating to 613 K for 4 times and isothermal heating at 613, 633, 653, and 673 K for 10 min. (c) Raman spectra of ZIF-4, HAD glass, HPHDA, and $A_{593\text{-}30}$HPHDA. (d) Faber-Ziman structure factors *S*(Q) of ZIF-4, HDA glass, HPHDA, and $A_{633\text{-}30}$HPHDA. (e) Reduced pair correlation functions *G*(*r*) of ZIF-4, HDA glass, HPHDA, and $A_{633\text{-}30}$HPHDA. (f) Schematic diagram of the atomic relationships in the ZIF-4 structure. Note: magenta spheres represent Zn, gray spheres represent C, cyan-green spheres represent N, and light pink spheres represent H.

## 2.2. Photoluminescence performance

Based on the observation of the strong fluorescence of HDA glass, HPHDA, and $A_{593\text{-}10}$HPHDA (Fig. 2c), it is interesting to study the significant impact of the subtle tetrahedral rearrangement on the electronic structure of the glasses, leading to photonic performance. Fig. S9 (Supporting Information) shows the optical absorption spectra of ZIF4, HDA glass, and HPHDA. ZIF-4 and HDA glass exhibit weak absorption in the visible range, but show a bit stronger absorption in the ultraviolet range (below 350 nm). In addition, the electronic structure of ZIF-4 is calculated using periodic density functional theory (PDFT) to gain insight into the PL mechanism. As shown in Fig. 3c, we observe a flat band electronic structure with a band gap of 5.02 eV. Based on the analysis of the total/partial density of states (DOS) (Fig. 3c), the electronic density is mainly determined by the *p*-orbitals of C and N atoms. Furthermore, the highest occupied molecular orbital (HOMO) and the lowest unoccupied molecular orbital (LUMO) are dominated by the *p*-orbitals of C and N in the Im rings (Fig. 10, Supporting Information). This indicates that the photoluminescence (PL) of the ZIF-4 crystal is primarily attributed to the $\pi$-$\pi^*$ interaction within the rings. However, a sharp increase in absorbance with decreasing wavelength is observed at around 400 nm in HPHDA.

Figs. 3a-d show the two-dimensional (excitation-emission) PL spectra of ZIF-4, HDA, HPHDA, and $A_{593\text{-}10}$HPHDA, respectively. A strong narrow emission peak is observed at 310 nm in the crystal under excitation in the wavelength range of 290 nm (Fig 2g), whereas a weak broad emission peak is seen in the visible range for HDA, HPHDA, and $A_{593\text{-}10}$HPHDA. Under excitation at 270 nm, HDA glass exhibits a narrow emission peak at 290 nm and a weak broad emission peak at 431 nm (Fig. 3f). Interestingly, compared to the ZIF-4 crystal, the narrow and the broad emission peaks in HDA glass and HPHDA shift with excitation wavelength towards shorter wavelengths, i.e., leading to a blue shift[21]. Under excitation at 310 nm, the narrow emission peak at 353 nm in ZIF-4 becomes weak or disappears upon vitrification (Fig. 3h). Interestingly, the weak emission at around 443 nm in ZIF-4 is strongly enhanced through vitrification, hot-pressing, and annealing after HP (Fig. 3i). Moreover, this emission peak shows a blue shift upon vitrification and hot-pressing, and exhibits a red shift upon annealing after hot-pressing. The structural origin of the observed PL behaviors can be explained as follows. First, the narrow emission peaks at 310 and 353 nm in the crystal appear only when Im rings are periodically arranged, which are attributed to their $\pi^*$-$\pi$ transitions in Im rings. Second, the blue shift of narrow emission peaks from 310 to 290 nm and the vanishing of the emission peak at 290 nm may be ascribed to the increased orientational disorder of the Im linkers and distortion of the local Zn-N coordination environment[22]. Third, the red shift could be attributed to the annealing-induced increase of ligand-to-ligand charge transfer (LLCT), resulting in increased conjugation area and therefore lowering of the energy level of the antibonding $\pi^*$ electron orbital[14].

The chromaticity coordinates of ZIF-4, HDA, HPHDA, and $A_{593-10}$HPHDA under 270, 290, 310, and 370 nm excitation are marked in CIE 1931 color spaces (Fig. S11, Supporting Information). The color point of $A_{593-10}$HPHDA lies in the light-blue lighting region, whereas those of ZIF-4, HDA, and HPHDA are located in the blue lighting region. Fig. 3j shows the absolute internal (black column) and external (red column) photoluminescence quantum yield (PLQY) values for ZIF-4, HDA, HPHDA, and $A_{593-10}$HPHDA under excitation at 365 nm. The internal PLQY of the ZIF-4 crystal under excitation at 365 nm first increases from 3.8 to 9.3% upon vitrification and then decreases to 5.2% by hot-pressing and finally reaches 8.6% upon annealing at 593 K for 30 min. The external PLQY of the ZIF-4 crystal follows the same trend as the internal PLQY, but with different absolute values. The fluorescence lifetime for ZIF-4 at 435 nm under excitation at 375 nm (Fig. 3k) is shortened by vitrification and hot-pressing, and then is extended by annealing at 593 K for 30 min.

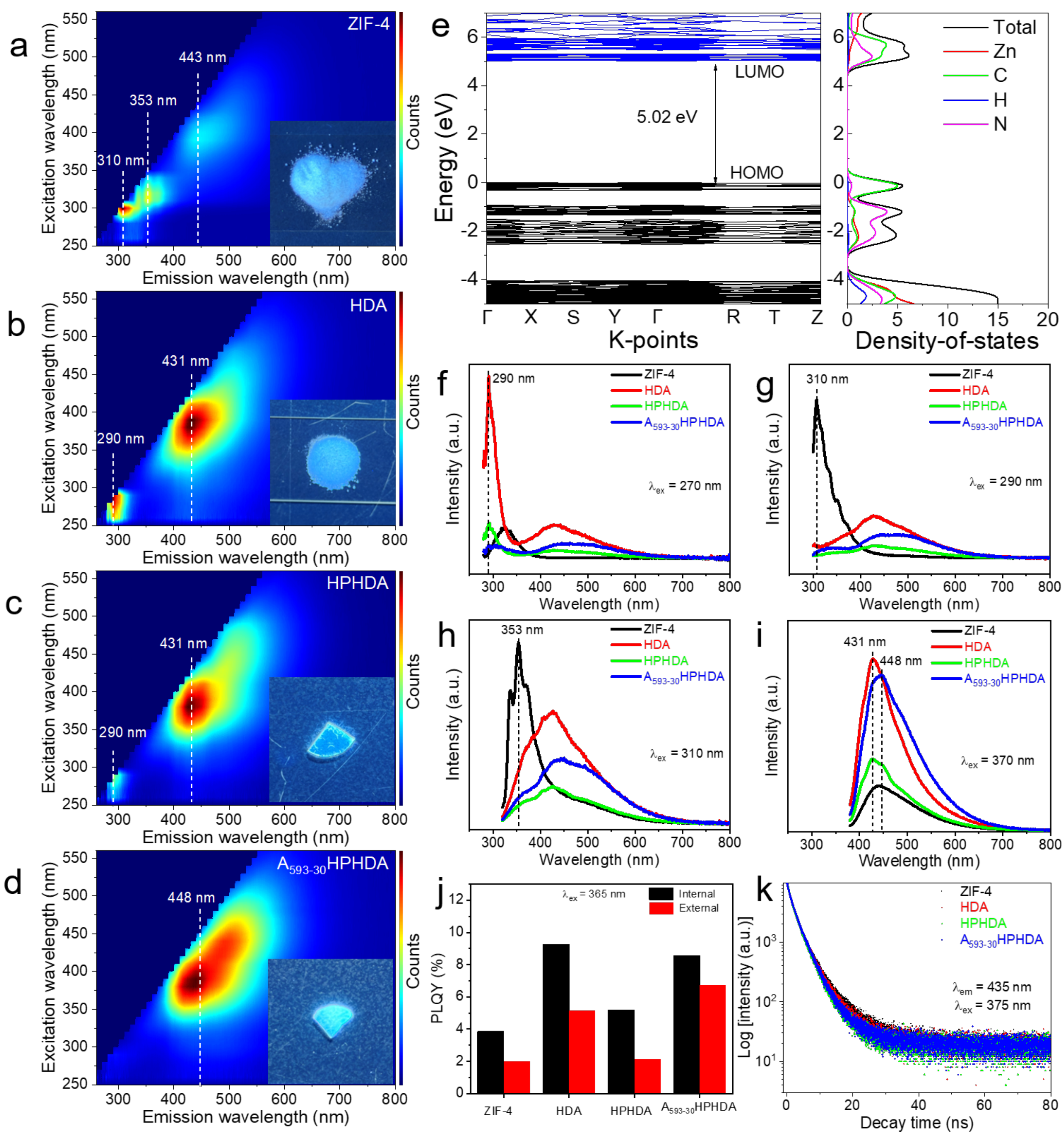

**Fig. 3 Photoluminescence (PL) performance of ZIF-4 crystal and glasses.** (a-d) Two-dimensional fluorescence (Excitation/Emission) spectra of (a) ZIF-4, (b) HDA, (c) HPHDA, and (d) $A_{593-30}$HPHDA. Inset: Optical photos of (a) ZIF-4, (b) HDA, (c) HPHDA, and (d) $A_{593-30}$HPHDA under a 365 nm laser. (b) Band structure and corresponding total/partial density of states (TDOS/PDOS) of ZIF-4. (e-h) Comparison of emission spectra of ZIF-4, HDA, HPHDA, and $A_{593-30}$HPHDA excited at 270 (f), 290 (g), 310 (h), and 370 nm (i), respectively. (j) Absolute internal and external PLQY of ZIF-4, HDA, HPHDA, and $A_{593-30}$HPHDA under excitation at 365 nm. (K) Room-temperature lifetime of ZIF-4, HDA, HPHDA, and $A_{593-30}$HPHDA samples under excitation at 375 nm.

## 2.3. Gas sorption performance

To evaluate the pore structure of ZIF-4 upon vitrification, hot-pressing, and annealing, the gas adsorption properties were preliminarily investigated. Based on Henke's report on $N_2$ and Ar physisorption for ZIF-4 and ZIF-62 samples[23], the smaller kinetic diameter of $CO_2$ (3.30 Å) relative to those of $N_2$ (3.64 Å) and Ar (3.40 Å), combined with the higher adsorption temperature (195 vs. 77 K), facilitates $CO_2$ diffusion into narrow micropores (<5 Å)[23][24][25]. Compared to the $CO_2$ sorption experiments performed at 273 or 298 K, measurements at 195 K enable complete filling of the micropores at 100 kPa. At 195 K, $CO_2$ would normally solidify in nonporous or macroporous systems. However, within micropores (≤20 Å), nanoscale confinement suppresses crystallization, allowing the adsorbed $CO_2$ to remain in a supercooled liquid state. Consequently, the micropores can be effectively filled with fluid $CO_2$, enabling the accessible micropore volume to be determined from the $CO_2$ uptake. Consequently, the maximum gas capacities ($n_{ads}$) determined at 95 kPa can be used to determine the specific micropore volume ($V_{pore}$) of the investigated ZIF-4 and ZIF-62 samples by taking into account the density of supercooled liquid $CO_2$ at 195 K (Figs. 4a and b). Therefore, the porosity of ZIF-4, HDA, HPHDA, $A_{633-10}$HPHDA, and $A_{773-5}$HDA was probed by $CO_2$ physisorption at 195K. Remarkably, all these phases adsorb $CO_2$ as evident from the Langmuir-shaped isotherms (Figs. 4a and b)

The $CO_2$ adsorption isotherms reveal that ZIF-4 crystals exhibit the highest $V_{pore}$ of 0.275 $cm^3 \cdot g^{-1}$, whereas vitrification into HDA glass drastically reduces $V_{pore}$ to 0.112 $cm^3 \cdot g^{-1}$. This pronounced decrease reflects substantial framework collapse and densification; nevertheless, approximately 41% of the micropore volume remains accessible in the HDA glass. In addition, upon hot-pressing and annealing at both 633 K for 10 min and 773 K for 5 min, the $V_{pore}$ of HDA glass reduces to 0.093, 0.092, and 0.069 $cm^3 \cdot g^{-1}$, respectively. Based on XRD and TEM results shown in Figs. 1 b-d, both the size of ZIF-zni crystals and the crystallinity of HDA glass increase with increasing annealing temperature. Notably, ZIF-zni is the densest and most stable $Zn(Im)_2$ crystal, indicating that the micropore size becomes small upon annealing.

Compared to ZIF-4, the micropore volume of crystalline ZIF-62 decreases to 0.201 $cm^3 \cdot g^{-1}$. This correlates with the implementation of the secondary bulky imidazolate linkers (bIm) in ZIF-62, reducing the void space of the crystalline framework. The pore volume of the melt-quenched glass (MQG), derived from a ZIF-62 crystal, is 0.137 $cm^3 \cdot g^{-1}$, showing a reduction of 25% compared to that of the ZIF-62 crystal. This observation suggests that MQG has a more similar pore structure to that of its crystalline ZIF-62, while HDA glass differs strongly from crystalline ZIF-4. Interestingly, the $V_{pore}$ of MQG reduces to 0.110 $cm^3 \cdot g^{-1}$ upon hot-pressing, and increases to 0.148 $cm^3 \cdot g^{-1}$ upon annealing at 653 K for 10 min. Moreover, direct annealing of MQG at 653 K for 10 min induces an increase in $V_{pore}$ from 0.137 to 0.142 $cm^3 \cdot g^{-1}$, which is different from that in HPHDA glass. This significant improvement in $CO_2$ adsorption for ZIF-62 glass suggests that the relaxation of the distorted structure towards a less distorted structure upon annealing above $T_g$. This relaxation leads to annealed MQG and HPG staying at a lower potential energy level state[14]. Consequently, band-gap energy ($E_g$) between the valence band and the conduction band decreases upon annealing, resulting in a slight red shift in the PL of both HPG and MQG. For the above reasons, the mechanism of the PL red shift upon annealing above $T_g$ needs to be further investigated. However, it is important to note that the pore structure have an srong impact on the position of PL. Moreover, the mechanical performance of the studied ZIF glasses remains inferior to that of conventional oxide glasses[26][27], highlighting the need for further efforts to enhance their mechanical properties[28].

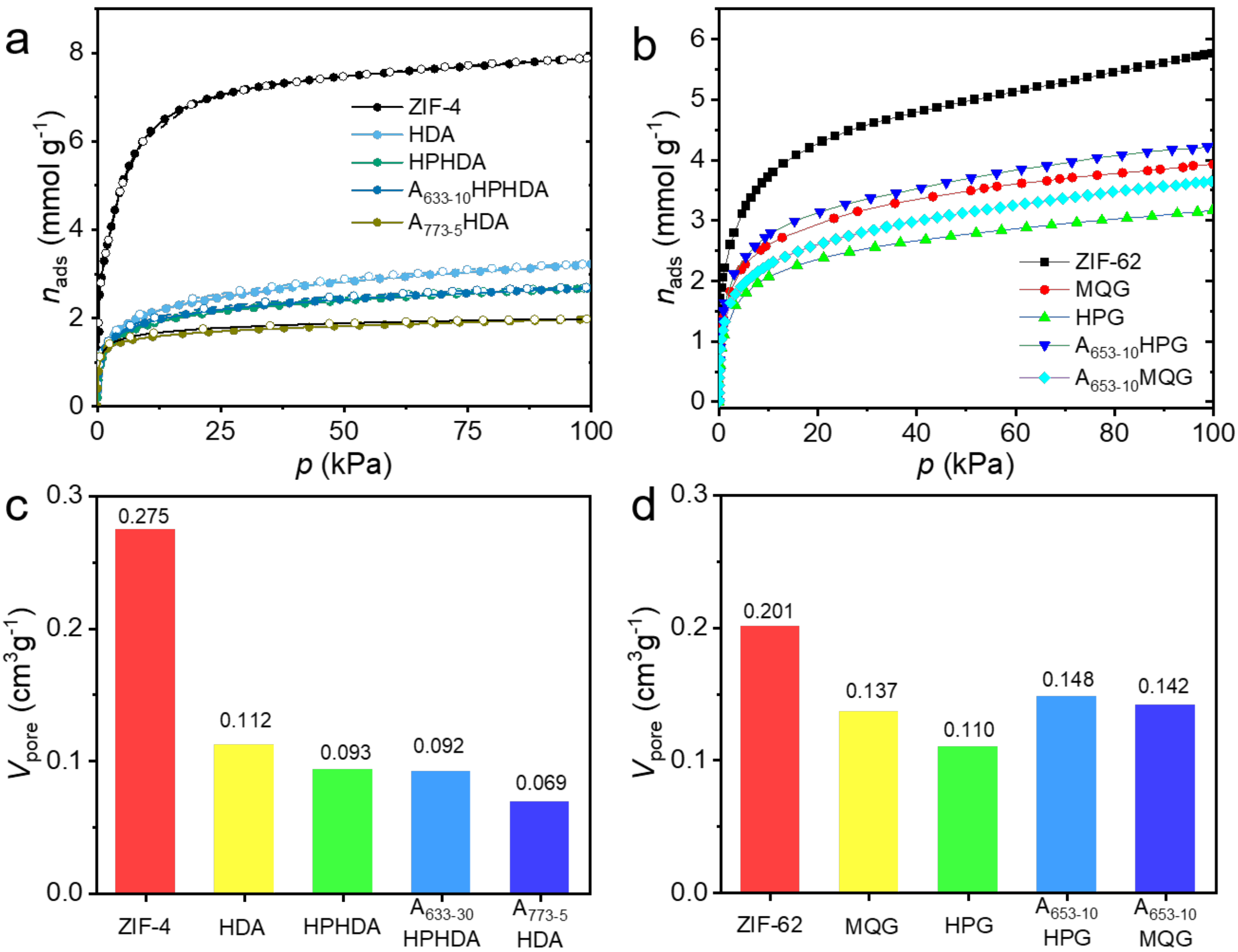

**Fig. 4 $CO_2$ gas sorption analysis of ZIF-4 and ZIF-62 crystals and glasses.** (a) $CO_2$ sorption isotherms collected at 195 K of ZIF-4, HDA, HPHDA, $A_{633\text{-}10}$HPHDA, and $A_{773\text{-}5}$HDA. (b) $CO_2$ sorption isotherms collected at 195 K of ZIF-62, MGQ, HPG, $A_{653\text{-}10}$HPG, and $A_{653\text{-}10}$MQG. (c) Bar plot of the micropore volumes ($V_{pore}$) for ZIF-4, HDA, HPHDA, $A_{633\text{-}10}$HPHDA, and $A_{773\text{-}5}$HDA derived from the $CO_2$ isotherms at 195 K. (d) Bar plot of the micropore volumes ($V_{pore}$) for all crystalline ZIF-62 and their corresponding glasses derived from the $CO_2$ isotherms at 195 K.

## 3. Conclusions

We synthesized ZIF-4 crystals via the solvothermal approach and prepared HDA glasses through quenching high-density liquid, hot-pressing, and annealing. We discovered the narrowband ultraviolet emission and the weak broadband visible light emission from the ZIF-4 crystal. More importantly, the broadband visible light emission was enhanced by means of vitrification, hot-pressing, and annealing, while a significant red shift was induced by annealing at a temperature above $T_g$. These phenomena were attributed to the band gap decrease caused by structure relaxation. Such structural evaluation was verified by high-energy synchrotron X-ray diffraction measurements. Based on gas adsorption results, we found that the $V_{pore}$ of MQG reduces to 0.110 $cm^3 \cdot g^{-1}$ upon hot-pressing, and increases to 0.148 $cm^3 \cdot g^{-1}$ upon annealing at 653 K for 10 min. Moreover, direct annealing of MQG at 653 K for 10 min induces an increase in $V_{pore}$ from 0.137 to 0.142 $cm^3 \cdot g^{-1}$, which is different from that in HPHDA glass. This work provides important insights into the porosity of MOF glasses and thus contributes to the understanding of the structure of such glasses. Future advancements, such as new design concepts for tuning and adjusting the porosity and sorption selectivity of MOF glasses, must be developed.

## Acknowledgments

The authors acknowledge the support from the i15 beamline (Proposal No. CY39002) at Diamond Light Source (DLS) in Didcot, United Kingdom. The authors also acknowledge the Carlsberg Foundation (Grant No. CF21-0371) for funding the STA 449 F3 DSC used in this work.

## Conflict of Interest

The authors declare no conflicts of interest.